\documentclass[useAMS,usenatbib,a4paper]{mn2e}

\usepackage{natbib,amsmath,amssymb}
\usepackage{epsfig,graphicx,xspace}
\usepackage[colorlinks=true,linkcolor=blue,citecolor=blue]{hyperref}
\usepackage{breakurl}
\usepackage{mathrsfs}

\def\aap{A\&A}%
\def\aapr{A\&ARv}%
\def\apj{ApJ}%
\def\apjl{ApJ}%
\def\mnras{MNRAS}%
\def\nat{Nature}%
\def\araa{ARA\&A}%
\def\prd{Phys.~Rev.~D}%

\def\apss{Ap\&SS}
\def\aj{AJ}
\def\jcap{JCAP}

\usepackage[usenames,dvipsnames,svgnames,table]{xcolor}

\title[The $\gamma$-ray afterglows of TDEs] {The Gamma-ray Afterglows of Tidal Disruption Events}
\author[Xian Chen et al.]{Xian Chen$^{1}$\thanks{E-mail: xchen@astro.puc.cl}, Germ\'{a}n G\'{o}mez-Vargas$^{1,2,3}$ \& James Guillochon$^{4,5}$\\
$^1$Instituto de Astrof\'isica, Facultad de F\'isica, Pontificia Universidad Cat\'olica de Chile, 782-0436 Santiago, Chile\\
$^2$Instituto de F\'isica, Pontificia Universidad Cat\'olica de Chile, Avenida Vicu\~na Mackenna 4860, Santiago, Chile\\
$^3$Istituto Nazionale di Fisica Nucleare, Sezione di Roma ``Tor Vergata'', I-00133 Roma, Italy\\
$^4$Harvard-Smithsonian Center for Astrophysics, The Institute for Theory and
Computation,\\
~~60 Garden Street, Cambridge, MA 02138, USA\\
$^5$Einstein Fellow}

\begin{document}

\date{Draft \today}

\pagerange{\pageref{firstpage}--\pageref{lastpage}} \pubyear{2015}
\maketitle

\label{firstpage}

\begin{abstract}
A star wandering too close to a supermassive black hole (SMBH) will be tidally
disrupted. Previous studies of such ``tidal disruption event'' (TDE) mostly
focus on the stellar debris that are bound to the system, because they give
rise to luminous flares. On the other hand, half of the stellar debris in
principle are unbound and can stream to a great distance, but so far there is no
clear evidence that this ``unbound debris stream'' (UDS) exists. Motivated by
the fact that the circum-nuclear region around SMBHs is usually filled with
dense molecular clouds (MCs), here we investigate the observational signatures
resulting from the collision between an UDS and a MC, which is likely to happen
hundreds of years after a TDE. We focus on $\gamma$-ray emission ($0.1-10^5$
GeV), which comes from the encounter of shock-accelerated cosmic rays with
background protons and, more importantly, is not subject to extinction.  We
show that because of the high proton density inside a MC, the peak $\gamma$-ray
luminosity, about $10^{39}~{\rm erg~s^{-1}}$, is at least $100$ times greater
than that in the case without a MC (only with a smooth interstellar medium).
The luminosity decays on a time-scale of decades, depending on the distance of
the MC, and about a dozen of these ``TDE afterglows'' could be detected within
a distance of about $16$ Mpc by the future Cherenkov Telescope Array.  Without
careful discrimination, these sources potentially could contaminate the
searches for starburt galaxies, galactic nuclei containing millisecond pulsars
or dark-matter annihilation signals.
\end{abstract}


\section{Introduction}\label{sec:intro}

A star wandering too close to a supermassive black hole (SMBH, with mass
$M_\bullet\ga10^6~M_\odot$) would be tidally disrupted, an incident known as
the ``tidal disruption event'' \citep[TDE,][]{hills75,rees88}.  The disruptive
process starts when the tidal force exerted by the SMBH becomes comparable to
the self-gravity of the star. Given the mass $M_*$ and radius $R_*$ of the
star, this happens at a critical distance of $R_t\simeq
R_*(M_\bullet/M_*)^{1/3}$ from the SMBH \citep{hills75}. Since this ``tidal
radius'' is tiny -- it is merely $23~(M_\bullet/10^6M_*)^{-2/3}$ times greater
than the Schwarzschild radius of the SMBH -- the TDE rate in a galaxy is
usually low, about ${\cal O}(10^{-4})~{\rm yr^{-1}}$ according to theoretical
calculations \citep[e.g.][]{magorrian99,wang04}.

A fraction of the star after tidal disruption has a negative energy and remains
gravitationally bound to the SMBH.  Mutual collisions and the later accretion
(by SMBH) of these bound stellar debris are expected to give rise to an
electromagnetic outburst with a thermal spectrum, which peaks at UV and soft
X-ray bands \citep{rees88}. This ``tidal flare'' provides an effective means to
reveal an otherwise dormant SMBH. So far tens of TDEs have been discovered in
this way \citep[see][for a review]{komossa15}.  

In principle, about half of the debris released from the disrupted star would
gain a positive energy and escape from the system with a velocity
asymptotically approaching $10^3-10^4~{\rm km~s^{-1}}$ \citep{rees88}.
Simulations showed that these unbound material will develop into an elongated
``unbound debris stream''
\citep[UDS,][]{kochanek94,guillochon09,guillochon15b,coughlin15}.  Two
observational signatures directly associated with the UDS have been envisaged
in the literature: (i) A brightening of Hydrogen lines several days after a TDE
caused by the recombination of the initially ionized plasma of the UDS
\citep{kasen10}, and (ii) variability of optical emission lines  on a
time-scale of months to years due to the illumination of the UDS by the central
tidal flare \citep{strubbe09,strubbe11}. While broad lines have been detected
from TDEs \citep{gezari12}, the detected lines likely originate from the
circularizing disk of debris \citep{guillochon14} rather than the two
aforementioned UDS signatures.

As a result, one important element in the current TDE model -- that at least
half of the stellar mass becomes unbound -- remains untested. A possible cause
of the non-detection is attenuation of optical/UV photons, by either the
interstellar medium (ISM) close to SMBHs \citep{donley02,gezari09} or a gaseous
envelope shrouding the radiative debris
\citep{loeb97,ulmer99,strubbe09,metzger15,miller15}. For this reason, it is
important to look for signatures of UDSs in an electromagnetic waveband that is
less subject to extinction, such as radio, infrared, hard X-ray or $\gamma$-ray
bands.

Such a signature potentially could be generated by the collision of an UDS
with the surrounding ISM.  It has been realized since about two decades ago
\citep{khokhlov96} that the UDS-ISM interaction will create a structure similar
to a supernova remnant (SNR). Recently, we started investigating the radiation
from this ``unbound debris remnant'' \citep[UDR,][]{guillochon15b}. We found
that if it exists in an environment like the Galactic Centre (GC), the spectrum
would peak in UV and soft X-ray, where extinction is still an issue.  However,
if a significant fraction of the collisional energy ends up in accelerating
electrons, radio emission will be produced due to synchrotron radiation and the
luminosity could amount to ${\cal O}(10^{37})~{\rm erg~s^{-1}}$ (assuming a TDE
rate of $10^{-4}~{\rm yr^{-1}}$).  Given this luminosity, it is possible to
detect ${\cal O}(1)$ such remnant in the GC, but not in other galaxies because
the object would be too faint \citep{guillochon15b}.

In our earlier studies of the UDR, we had not considered radiative mechanisms
of hard photons, such as hard X-ray and $\gamma$-ray.  It is worth mentioning
that one way of generating hard X-rays/$\gamma$-rays is by forming a jet, which
can up-scatter low-energy photons to higher energies by inverse Compton
process.  This mechanism was introduced to explain the transient hard X-ray
($15-50$ keV) emission detected by {\it Swift} in three TDEs
\citep{bloom11,burrows11,levan11,zauderer11,cenko12,brown15}. However, jets are
not directly related to UDSs, therefore, jetted TDEs cannot be used for our
purpose of looking for UDSs, not to mention the fact that the
hard-X-ray/$\gamma$-ray radiation from a jet is highly beamed so that the
probability of seeing it is small. 

Recently, an alternative way of producing $\gamma$-ray emission was proposed,
originally to explain the possible excess of GeV $\gamma$-rays in the GC
\citep{cheng06,cheng07,cheng11,cheng12}. This type of $\gamma$-ray emission is
closely linked to the UDS-ISM collision. The collision basically generates a
large amount of relativistic protons, known as cosmic rays (CRs), through a
process called diffusive shock acceleration \citep{hinton09,treumann09}. The
$\gamma$-ray emission is produced when the CRs escape from the shock region and
start bombarding the non-relativistic protons in the ISM, leading to the
formation of neutral pions ($\pi^0$) which almost immediately decay into
$\gamma$-rays \citep[$\ga10$ MeV,][]{kafexhiu14}.

According to the calculations done for the GC \citep{cheng06,cheng07}, the
$\gamma$-ray luminosity would sustain at a level of about $10^{38}~{\rm
erg~s^{-1}}$ if the TDE rate is $10^{-5}~{\rm yr^{-1}}$. Given this luminosity,
a $\gamma$-ray telescope whose sensitivity is typically $10^{-13}~{\rm
erg~cm^{-2}~s^{-1}}$ at present \citep{funk13} is able to detect the above
source out to a distance of about $3$ Mpc.
 
It is a good time to revisit this latter mechanism of $\gamma$-ray emission
because of two instances of recent progress. (i) Excess of $\gamma$-ray
($0.1-100$ GeV) emission has been clearly detected in the GC \citep{fermi15},
and possibly in two nearby dwarf galaxies (reported by
\citealt{GS15,hooper15,li15}, but see \citealt{drlica-wagner15} for a different
result), by the Fermi Large Area Telescope (Fermi-LAT) on-board the Fermi
Gamma-ray satellite.  It is important to understand the contribution to these
$\gamma$-ray emission by TDEs.  (ii) Both theoretical and observational studies
(see \S\ref{sec:MC} for details) suggest that around SMBHs the ISM is
likely very clumpy, filled with long-lived molecular clouds (MCs). If UDSs
collide with MCs, an enhancement of $\gamma$-ray emission is anticipated
because the high proton density inside MC accelerates the production and the
subsequent decay rates of CRs.

Motivated by the latest progress,  we will analyze in this paper the
$\gamma$-ray emission from the UDS-MC collision and investigate the prospect of
detecting them in other galaxies.  The paper is organized as follows. In
\S\ref{sec:MC} we briefly review the evidences supporting the ubiquity of MCs
around SMBHs, and then we summarize the typical properties of such MCs.  Then
in \S\ref{sec:initial} we describe the characteristics of a typical UDS, so
that in \S\ref{sec:shock} we can use them to evaluate the time-scales and the
energetics related to the UDS-MC collision.  Based on these results, we discuss
in \S\ref{sec:cr} the properties of the CRs produced by the collision.  In
\S\ref{sec:gammaray} we calculate the luminosity and spectral energy
distribution of the $\gamma$-ray emission resulting from the cooling of the
above CRs inside the MCs of our interest.  Using these results, we evaluate in
\S\ref{sec:detect} the number of point sources that can be detected by
$\gamma$-ray telescopes.  Three telescopes are considered here, namely,
Fermi-LAT \citep{atwood09},  the High Energy Stereoscopic System
\citep[H.E.S.S., ][]{aharonian06} and the Cherenkov Telescope Array (CTA)
currently under planning \citep{acharya13}.  For completeness, in
\S\ref{sec:noMC} we also calculate the $\gamma$-ray emission resulting from the
interaction of UDSs with a smooth ISM, from which we infer the additional
number of detectable sources.  Finally, in \S\ref{sec:dis} we discuss the
possibility of detecting similar $\gamma$-ray sources in dwarf galaxies, due to
the existence of intermediate massive black holes (IMBHs), and we also outline
possible methods of separating our objects from other $\gamma$-ray sources.

\section{Molecular Clouds}\label{sec:MC}

There are at least three pieces of evidence suggesting that the ISM around
SMBHs is very clumpy.

(i) It is now widely accepted that SMBHs grew mainly during the phase of active
galactic nucleus (AGN). In the unified model of type-I/type-II AGNs,  the most
essential ingredient is a torus-like structure of several parsecs in size
composed of dusty gas clumps surrounding the central SMBH
\citep{krolik88,antonucci93}.  Given the ubiquity of this structure in AGNs, it
is natural to suspect that many SMBHs still have relic tori around them even
though they are no longer active today. 

(ii) Another evidence comes from a special type of galaxies only recently
discovered \citep{komossa08,komossa09,wang12,yang13}.  These galaxies show
peculiar iron and oxygen emission lines, which are atypical of a normal AGN but
more consistent with a model in which an ionizing spectrum similar to that of a
TDE is reflected by dense MCs \citep[e.g.][]{wang12}.  Besides, the line
strength decays on a time-scale of several years, indicating that the
reflectors reside within several parsecs from the central SMBHs
\citep[e.g.][]{komossa08}.

(iii) The most direct evidence is from the GC. Radio observations revealed tens
of MCs at a distance of $0.5-2$ pc from Sgr A$^*$, the SMBH in the GC
\citep{mezger96}. The densities of these MCs lie in a wide range of
$n_H\sim(10^6-10^8)~{\rm cm^{-3}}$ \citep{christopher05}, similar to what has
been inferred for the MCs in those galaxies with peculiar line ratios
\citep{wang12}.  The spatial distribution and kinematics of the MCs in the GC
are suggestive of a torus-like structure with a vertical thickness of about
$0.5$ pc at the inner edge \citep{christopher05}.  Given these observed
properties, the torus has a half-opening angle of about $27^\circ$ and covers
in total $f_c\sim40\%$ of the sky when viewed from the central SMBH.  It has
been pointed out that such a geometry resembles that of a normal AGN torus
\citep{mezger96,ponti13}.

Based on these evidences, it seems likely that after a TDE, the UDS would
collide with one of the MCs around the central SMBH.  This collision differs in
two ways from the interaction with a smooth ISM.  (i) Because of the much
higher proton density inside MC, the UDS loses its kinetic energy more rapidly
\citep{guillochon15b}, which implies a higher production rate of CRs. (ii) The
CRs, after escaping from the shock region, are still inside a very dense
environment (the MC), a condition favourable to $\pi^0$ formation.  The second
difference is also the main reason that those MCs close to supernova remnants
(SNRs) often show an enhanced $\gamma$-ray emission
\citep{aharonian96,fatuzzo06,gabici15,hess15}.

It is important to note that the MCs close to SMBHs are not the same as those
seen in the Galactic plane. The former are orders of magnitude denser than the
latter, and the reason is as follows.  Given a typical distance of $D\sim1$ pc
between our MC and a SMBH, the Roche limit, which is a criterion for the MC to
remain gravitationally bound, requires that
\begin{equation}
n_H\ga M_\bullet/(m_pD^3)\simeq4.1\times10^7m_6D_1^{-3}~{\rm cm^{-3}},\label{eq:nH}
\end{equation}
where $m_p$ is the proton mass, $m_6=M_\bullet/(10^6~M_\odot)$ and
$D_1=D/(1~{\rm pc})$.  Therefore, when $D\sim1$ pc and
$M_\bullet\ga10^6~M_\odot$, we have $n_H\ga4\times10^7~{\rm cm^{-3}}$. This
density is orders of magnitude higher than most MCs in the Galactic plane
[$n_H\sim(10^2-10^3)~{\rm cm^{-3}}$], and in fact even denser than the cores of
normal MCs \citep[$n_H\sim10^5~{\rm cm^{-3}}$,][]{bodenheimer11}.

Despite such a high density, the MCs around SMBHs do not collapse to form stars
(Jeans instability), probably because they are not self-gravitating but
confined mainly by an external agent, such as the surrounding hot and turbulent
ISM, or the compressive tides of the central SMBHs
\citep[e.g.][]{chandra63,shu92}.  In equilibrium with an external force, a MC
could have a density significantly lower than what is required by the Roche
limit \citep{chen15b}.  For this reason, we chose $n_H=10^7~{\rm cm^{-3}}$ as
our fiducial value. This value agrees within a factor of $2-3$ with the mean
density $4\times10^6~{\rm cm^{-3}}$ of those MCs in the GC, inferred from the
brightness of molecular emission lines \citep[optical-thin density
from][]{christopher05}.

To complete our description of a MC, we take $R_c=0.25$ pc as the fiducial
radius, which is typical of the MCs in the GC as well \citep{christopher05}.
Given this size and assuming that our MC is spherical and homogeneous, we can
derive a mass of $M_c=4\pi m_pn_HR_c^3/3\sim1.6\times10^4n_7~M_\odot$ for the
cloud, where $n_7\equiv n_H/(10^7{\rm cm^{-3}})$.

\section{Unbound debris streams}\label{sec:initial}

The evolution of an UDS in vacuum constitutes an important part of the initial
conditions of our problem. Since a series of earlier works have studied this
topic \citep{rees88,kochanek94,khokhlov96,strubbe09,kasen10}, we only summarize
here the main results that are most relevant to our work, including the kinetic
energy, internal velocity dispersion and the opening angle with respect to the
central SMBH.

An UDS originates from the part of a star which at the time of tidal disruption
has a positive energy.  This unbound part contains about half of the mass of
the star if the star initially approaches the SMBH along a parabolic orbit with
a periapsis distance of $R_t$ -- an probable case in real galaxies
\citep{rees88,magorrian99}. In this case, the unbound material distribute in a
range of specific energy between ${\cal E}=0$ and ${\cal E}={\cal
E}_0=GM_\bullet R_*/R_t^2$ \citep{guillochon13,stone13}. 
As a result, the total kinetic energy $E_k$ of the UDS is
\begin{equation}
{E}_k  =\int_0^{{\cal E}_0}{\cal E}\frac{dM}{d{\cal E}}d{\cal E}
\simeq \frac{M_*{\cal E}_0}{4}\simeq 10^{50}~{\rm
erg}~m_6^{1/3}m_*^{5/3}r_*^{-1},\label{eq:Ek}
\end{equation}
where $m_*\equiv M_*/M_\odot$, $r_*\equiv R_*/R_\odot$. Here we have assumed an
even distribution of unbound material in specific energy, i.e.  $dM/d{\cal
E}=M_*/(2{\cal E}_0)$, which was proposed by \cite{rees88} and later confirmed
by smooth-particle hydrodynamics simulations \citep{evans89}. 

Although we made a couple simplifications to derive Equation~(\ref{eq:Ek}), the
result agrees well with the median value computed from a statistically
representative ensemble of UDSs \citep{guillochon15b}.  Using this equation and
the fact that $r_*\propto m_*^{0.6}$ for main-sequence stars at $m_*>1$
\citep{demircan91,gorda98}, we further derive $E_k\propto m_*^{1.1}$. This
scaling relation indicates that more massive stars generally give rise to more
energetic UDSs. 

It is worth noting that Equation~(\ref{eq:Ek}) does not include the additional
gain of kinetic energy during the compression and bouncing of the star at the
orbital periapsis \citep[][and references therein]{brassart08}. We neglected it
because the bouncing energy ${\cal E}_b$, about $\beta^2GM_*/R_*$ per unit mass
\citep{guillochon09}, is usually much smaller than ${\cal E}_0$, where
$\beta\equiv R_t/R_p$ is the so-called ``penetration factor'' defined as the
ratio between $R_t$ and the pericentre distance $R_p$ of the initial stellar
orbit.

Take tidal disruption of a solar-type star ($M_*=M_\odot$ and $R_*=R_\odot$)
for example.  When $M_\bullet=4.3\times10^6~M_\odot$ which is the mass of the
SMBH in the Milky Way (MW) \citep{genzel10}, the penetration factor $\beta$
cannot be greater than $4.3$; otherwise the star will enter a radius smaller
than the ``innermost bound circular orbit'' (twice the Schwarzschild radius for
a non-rotating black hole) and eventually be swallowed as a whole by the SMBH.
Because of such a limit to $\beta$, the ratio ${\cal E}_b/{\cal
E}_0\simeq\beta^2(M_*/M_\bullet)^{1/3}$ is small, in fact in the range
$(0.0063,0.12)$ for the SMBHs in the mass range of
$10^6<M_\bullet/M_\odot<10^8$.

We notice that our ${E}_k$ differs by a factor of $\beta^2$ from that derived
by \citet{khokhlov96,strubbe09,kasen10}. This is because those authors
calculated ${\cal E}_0$ using the formula $GM_\bullet R_*/R_p^2$, i.e. they
assumed that the energy spread is determined at $R_p$ instead of at $R_t$.
However, recent studies of the tidal-disruption process suggest that using
$R_t$ is physically more appropriate \citep{guillochon13,stone13}.  Therefore,
using $R_p$ in the calculation will result in an overestimation of ${\cal E}_0$
by a factor of $\beta^{2}$, which is non-negligible when $\beta>1$.

Besides $E_k$, another important quantity of an UDS is its velocity. It
determines the time of collision with a MC as well as the Mach number of the
subsequent shock. It is important to realize that different parts of an UDS
travel at different velocities because of the energy difference imprinted in
different parts of the star at the time of tidal disruption.  The fastest part
has the highest specific energy, which, as we already know, is ${\cal E}_0$. As
a result, it has the highest velocity. When it has traveled to a distance of $D\gg
R_t(M_\bullet/M_*)^{1/3}$, its velocity has asymptotically dropped to
\begin{equation} v_0\simeq\sqrt{2{\cal E}_0}\simeq6.2\times10^3~{\rm km~s^{-1}}~
m_6^{1/6}m_*^{1/3}r_*^{-1/2}. 
\end{equation}

Suppose the time $t=0$ coincides with the time of stellar disruption, then the
time at which the fastest part of the UDS has traversed a distance of $D$ is
\begin{equation} t_0\simeq D/v_0\simeq1.6\times10^2~{\rm
yr}~m_6^{-1/6}m_*^{-1/3}r_*^{1/2}D_1.\label{eq:t0} \end{equation}
For our interest $D\sim1~{\rm pc}\gg R_t(M_\bullet/M_*)^{1/3}$.  As for the
part with a lower specific energy, i.e. ${\cal E}<{\cal E}_0$, it follows that
the velocity asymptotically approaches $\sqrt{2{\cal E}}$ and it arrives at $D$
later, around the time $t\simeq D/\sqrt{2{\cal E}}$.

Not only the velocities are different,  the directions where different parts of
an UDS are headed also differ \citep[e.g.][]{kochanek94}.  In the equatorial
plane -- the plane defined by the initial orbit of the progenitor  star -- the
difference is caused mainly by the energy spread ${\cal E}_0$.  The
corresponding angular span relative to the central SMBH can amount to
$2v_0/v_p$ when the self-gravity of UDS is negligible \citep{khokhlov96}, where
$v_p\simeq\sqrt{2GM_\bullet/R_p}$ is the mean stellar velocity during the
pericentre passage.  In the direction perpendicular to the equatorial plane,
the angular span is caused by the bouncing velocity
$v_\perp\simeq\beta\sqrt{GM_*/R_*}$ and is about $2v_\perp/v_p$
\citep{kasen10}. From these angles, we derive a typical solid angle of
$\Omega\simeq\pi v_0v_\perp/v_p^2=\pi[M_*/(2M_\bullet)]^{1/2}$ for UDSs.

According to the last formula, if a solar-type star is tidally disrupted by a
SMBH with a mass of $M_\bullet=(10^6,\,10^8)~M_\odot$, we find that
$\Omega=(2.2\times10^{-3},\,2.2\times10^{-4})$.  These values are consistent
with those from the earlier studies of UDSs, which showed that  $\Omega$
generally lies in the range of $(10^{-4},10^{-2})$.  It is known that the
uncertainty is due partly to the different ways of calculating ${\cal E}_0$
(e.g.  \citealt{khokhlov96,strubbe09,kasen10} used $R_p$ to calculate ${\cal
E}_0$), and partly to the additional consideration of the effects of
self-gravity \citep{kochanek94,coughlin15}.

To account for these theoretical uncertainties, we parameterize $\Omega$ with
$\Omega=10^{-3}\Omega_{-3}$ and take $\Omega_{-3}=1$ as our fiducial value.  In
any case, we find that $\Omega$ is much smaller than the typical solid angle,
$\pi R_c^2/D^2\simeq0.20D_1^{-2}$, of those MCs considered in the previous
section. This result  indicates that an UDS interacts with only one MC at a
time. 

\section{Collision between UDS and MC}\label{sec:shock}

So far we have specified the initial conditions of MCs and UDSs. Now we can
proceed to study their collisions.  Such a collision happens mostly likely at a
distance of ${\cal O}(1)$ pc from a SMBH, where MCs exist in large amount
(\S\ref{sec:MC}). An UDS arriving at this distance is in a free-expansion
phase, because the amount of ISM that has been swept up by the UDS is too small
to affect the kinematics.  For example, if the ISM surrounding the SMBH has a
constant density of $10^4~{\rm cm^{-3}}$ \citep[as has been assumed
by][]{khokhlov96,cheng06}, the swept-up mass would be only
$0.08D_1^3\Omega_{-3}~M_\odot$.  As a result, the UDS behaves in the same way
as in vacuum until the time of collision, at about $t\simeq t_0$. 

Immediately after the collision the UDS still continue its free expansion. But
as it advances deeper into the dense MC, it soon sweeps up an amount of material
that is as massive as the stream itself. When this happens, the free-expansion
phase ends and a Sedov-like expansion follows. This phase transition occurs at
a depth of

\begin{equation}
\Delta D\sim\frac{M_{\rm ej}}{n_H m_p\Omega D^2}
\simeq0.0020~{\rm pc}~m_* n_7^{-1} \Omega_{-3}^{-1} D_1^{-2}\label{eq:DeltaD}
\end{equation}
from the surface of the MC, where $M_{\rm ej}=0.5M_*$ is the mass of the UDS.
The fact that $\Delta D\ll R_c$ suggests that the MC is thick enough to stop
the UDS completely. On the other hand, a normal MC like those in the Galactic
plane cannot stop a UDS, because the density, typically of $10^2-10^3~{\rm
cm^{-3}}$ \citep{bodenheimer11}, would be too low.

Since the collisional velocity initially is about $v_s\sim v_0$, which is
orders of magnitude higher than the typical (turbulent) sound speed
$c_s\sim20~{\rm km~s^{-1}}$ in our MC \citep{genzel10}, a strong shock will be
produced.  The shock efficiently dissipates the kinetic energy of the UDS,
converting it into heat. To estimate the heating rate, we first recall that
mass is injected into the collisional region at a rate of
\begin{equation} \frac{dM}{dt}=\frac{dM}{d{\cal E}}\left|\frac{d{\cal E}}{dt}\right|
\simeq\frac{M_*}{t_0}\left(\frac{t}{t_0}\right)^{-3},\label{eq:dMdt}
\end{equation}
where $t\ge t_0$. The steep dependence on $t$ implies that UDS is head-heavy:
From the arrival time of the fastest debris at $t=t_0$ to the time $1.4t_0$,
already half of the mass of the UDS ($50\%$ of $M_*/2$) has been deposited into
the MC, and in the following period of $0.6t_0$, much less matter ($25\%$ of
$M_*/2$) arrives. Knowing the mass-injection rate $dM/dt$, we can calculate the
energy-injection/heating rate with
\begin{equation}
\frac{d{E}_k}{dt}={\cal E}\frac{dM}{dt}
=\frac{4{E}_k}{t_0}\left(\frac{t}{t_0}\right)^{-5}.\label{eq:dEkdt}
\end{equation}
The even steeper dependence on $t$ indicates that kinetic energy is more
concentrated at the head of the UDS than is mass: From $t=t_0$ to about $1.2
t_0$, already $50\%$ of the total kinetic energy has been injected, and by
$t=2t_0$ the injection of kinetic energy is about $94\%$ complete.

To balance the heating rate, which is of order $E_k/t_0\sim2\times10^{40}~{\rm
erg s^{-1}}$ at the beginning of the collision,  the surface temperature of the
MC should rise to at least $T=80$ K, so that the cooling rate $4\pi \sigma_S
R_c^2T^4$ ($\sigma_S$ is the Stefan-Boltzmann constant) due to black-body
radiation could be higher than $2\times10^{40}~{\rm erg s^{-1}}$. We note that
most MCs detected in the GC do have a moderately high gas temperature, about
$50-200$ K \citep{genzel10}, therefore, it seems that cooling would be
effective.
 
Because of its ability of efficient cooling, our MC would not expand
significantly even though the total energy injected by the UDS, $E_k$, greatly
exceeds the gravitational binding energy of the cloud, which is about
${E}_c\sim GM_c^2/R_c\sim2.4\times10^{49}n_7^2~{\rm erg}$.  On the other hand,
in the (unlikely) case of insufficient cooling, the time-scale for our MC to
expand is about $R_c/c_s\sim2.4\times10^4$ yr, much longer than the collisional
time-scale $t_0$. Therefore, we can in any case neglect the dynamical evolution
of the MC during the collision, which significantly simplifies our later
analysis of the CR and $\gamma$-ray productions.  

Before moving on to the next section, it is worth mentioning that the material
immediately after the shock has an temperature of about $T_s\simeq
3m_pv_s^2/(16k_B)$ \citep[][$k_B$ is the Boltzmann constant]{inoue12,pan13},
which is of order of $10^8$ K for our system \citep[also see][]{guillochon15b}.
Such a hot medium would emit X-rays, but it would be difficult to see directly
this emission, because the MC of our interest has a column density of
$n_HR_c\sim6\times10^{24}n_7~{\rm cm^{-2}}$, i.e. it is Compton thick.
Consequently, the seed X-ray photons are likely absorbed inside the MC, and
what can be seen are the reprocessed, low-energy photons (because $T\simeq80$
K) emitted from the surface of the cloud.

\section{Cosmic rays}\label{sec:cr}

It is well known that a strong shock like what we have just described in the
previous section produces CRs.  The mechanism, known as the diffusive shock
acceleration, is well established due to the studies of SNRs \citep{hinton09}.
These earlier studies suggested that typically $\epsilon=10\%$ of the kinetic
energy injected into the shock region can be tapped to accelerate CRs, and
those CRs escaping from the shock region follow a power-law distribution in the
momentum space with an power-law index of $\gamma_0\simeq2$ \citep[the
``universal power law'', see review by][]{treumann09}. 

Applying these earlier results, we find that a total  amount of $\epsilon
E_k\sim10^{49}~{\rm erg}$ of CRs would be produced by the UDS-MC collision, and
in the space of kinetic energy $T_p$, which is a more convenient frame for the
later calculation of $\pi^0$ and $\gamma$-ray production, the CRs initially
follow a spectrum of
\begin{equation} 
dN_p/dT_p\propto(T_p+1)/(T_p^2+2T_p)^{(\gamma_0+1)/2}
\label{eq:CRspec} 
\end{equation}
\citep[also see][for the CR spectrum]{cheng06}.

If $\gamma_0\le2$, the CR spectrum must cut off at a maximum, $T_{p,{\rm
max}}$, to avoid divergence of the total kinetic energy.  This maximum energy
in principle depends on the power of the shock, the duration of the
particle-acceleration process and the strength of the magnetic field in the
shock region.  In practice, we will calculate $T_{p,{\rm max}}$ using the
formula $10^2 v_3^2t_2B_{\rm mG}$ TeV \citep{hinton09}, where
$v_3=v_s/(10^3~{\rm km~s^{-1}})$ characterizes the shock strength,
$t_2=t_s/(10^2~{\rm yr})$ is the shock-acceleration time-scale ($t_s$) in unit
of $100$ yr, and $B_{\rm mG}$ is the magnetic field strength in unit of mG.
This formula is derived under the assumption of Bohm diffusion, which is the
slowest diffusion process for particles to cross shock front.  Correspondingly,
it determines the lower limit of $T_{p,{\rm max}}$.

We now quantify the typical value of $T_{p,{\rm max}}$ in our model.  Since the
shock velocity initially is about $v_0$ and decays with time as $(t/t_0)^{-1}$
(\S\ref{sec:initial}), it follows that $v_s=(6000,\,3800)~{\rm km~s^{-1}}$ when
$t_s=t-t_0=(10,\,100)$ yr, where we have assumed $m_6=m_*=r_*=D_1=1$.  For
$B_{\rm mG}$, observations of the GC region showed that the magnetic field
varies from $2$ to $4$ mG around the edge of the MCs \citep[][]{YZ96}, and is
about $0.2$ mG in the field  between those MCs \citep[][]{crocker05}.
Therefore, we assume $B_{\rm mG}=1$, and finally we find that $T_{p,{\rm
max}}\simeq(0.36,\,1.4)$ PeV when $(t-t_0)=(10,\,100)$ yr. It is interesting to
note that the $T_{p,{\rm max}}$ resulting from the interaction between UDSs and
a smooth ISM also lies in the PeV range \citep{cheng12}.  Therefore, UDSs are
effective accelerators of PeV CRs. 

The CRs escaping from the shock region will bombard the non-relativistic
protons in the MC.  A significant fraction of the subsequent proton-proton
($pp$) collisions are inelastic, so that the CRs gradually lose their
kinetic energies and cool down.  The $pp$-collision time-scale can be
calculated with $(\sigma_{pp}n_H c)^{-1}$, where $\sigma_{pp}\simeq40$ mbarn is
the total collisional cross section and $c$ denotes the speed of light.  Taking
into account the inelasticity of the collision, usually parameterized by
$\kappa$ which has a value of about $0.45$ \citep{fatuzzo06}, we find a cooling
time-scale of
\begin{equation} 
\tau_{pp}=(\kappa\sigma_{pp}n_H c)^{-1}\simeq5.9~{\rm
yr}~n_7^{-1}.\label{eq:taupp} 
\end{equation}

During this period of $\tau_{pp}$, a CR would have diffused in the MC by a
length of $d\simeq ({\cal D}\tau_{pp})^{1/2}$, where ${\cal D}$ is the
diffusion coefficient. The typical value of ${\cal D}$ for those MCs in the
Galactic plane is about ${\cal D}({T_p})\simeq10^{26}({T_p}/10~{\rm
GeV})^{1/2}~{\rm cm^2~s^{-1}}$ \citep{ormes88,torres10}. However, for those in
the GC, the value of ${\cal D}$ is unclear except that it should be smaller
because of the stronger magnetic field. For our purpose, we will use
the diffusion coefficient of those normal MCs to derive an upper limit for $d$,
and we find that
\begin{equation}
d\la\sqrt{{\cal D}({T_p})\tau_{pp}}
\simeq0.25~{\rm pc}~n_7^{-1/2}\left[{T}_p/(10~{\rm TeV})\right]^{1/4}.
\label{eq:d}
\end{equation}

Now it is clear that $d< R_c$ when $T_p<10n_7^2$ TeV.  These inequalities mean
that a CR with a relatively low energy will be trapped inside MC because of
cooling.  In this case, the probability of $\pi^0$ production is the highest.
On the other hand, when $T_p\gg10n_7^2$ TeV, a CR may escape from the MC
(depending on ${\cal D}$, which is still unconstrained). As a result, the
kinetic energy of the CR is mostly lost, leading to a lower efficiency of
$\pi^0$ production. We will consider this effect in the later calculation of
the $\gamma$-ray spectrum resulting from $\pi^0$ decay.

So far we have neglected the energy loss of CRs due to collisional ionization.
This type of energy loss is relatively unimportant because the corresponding
time-scale, about $\tau_{\rm ion}\sim20~n_7^{-1}(T_p/1~{\rm GeV})$ years
\citep{berezinskii90}, is longer than $\tau_{pp}$, if we consider only the
energy range relevant to $\pi^0$ production which is $T_p\ga0.3$ GeV.

\section{Gamma rays}\label{sec:gammaray}

\subsection{Luminosity}\label{sec:L}

Having characterized the CRs, we can now investigate the production of $\pi^0$
by $pp$ collisions and calculate the subsequent $\gamma$-ray luminosity
$L_\gamma$.  We start from the formula \citep{fatuzzo06}
\begin{equation}
L_\gamma=\eta (\sigma_{pp}n_H c)E_{\rm CR}(t),\label{eq:Lgamma}
\end{equation}
where $E_{\rm CR}(t)$ is the total energy of CRs inside a MC and
$(\sigma_{pp}n_H c)$ is the collisional rate between a CR and the background
protons. The factor $\eta$ denotes the efficiency of $\pi^0$ production, and
when $\gamma_0$ increases from $2$ to $2.6$, $\eta$ decreases from $0.18$ to
$0.04$ \citep{crocker05}.

In Equation~(\ref{eq:Lgamma}) we deliberately wrote $E_{\rm CR}$ as a function
of $t$, so as to draw attention to its time dependence, which we now elaborate.
From the time $t=t_0$ when the UDS first hits the MC, to about $t_0+\tau_{pp}$,
the newly-produced CRs do not yet have time to cool down, so $E_{\rm CR}(t)$
increases monotonically. For this reason, we derive $E_{\rm CR}(t)$ by
integrating $\epsilon~dE_k/dt$, where $dE_k/dt\propto (t/t_0)^{-5}$ is the
energy injection rate coming from Equation~(\ref{eq:dEkdt}).  The result
is that $E_{\rm CR}(t)=\epsilon E_kC(t)$, where the function
$C(t)=1-(t/t_0)^{-4}$ comes from an integration of $(t/t_0)^{-5}$. Therefore,
we know that the luminosity evolves as
$L_\gamma=(\eta\epsilon/\kappa)E_kC(t)/\tau_{pp}$ when $t_0<t<t_0+\tau_{pp}$.

Afterwards ($t>t_0+\tau_{pp}$) cooling becomes relevant.  Consider now a time
interval $(t,t+\tau_{pp})$ in this later stage. During this interval, the CRs
produced earlier than $t$ would have cooled down completely and meanwhile a
fresh amount of $\epsilon(dE_k/dt)\tau_{pp}$ of CRs are created.  This
consideration suggests that $E_{\rm CR}(t)\sim \epsilon(dE_k/dt)\tau_{pp}$ at
any moment of $t>t_0+\tau_{pp}$.  Now using Equation~(\ref{eq:Lgamma}), we find
that the corresponding luminosity is
$L_\gamma\sim(\eta\epsilon/\kappa)~dE_k/dt\propto(t/t_0)^{-5}$.

\begin{figure}
\includegraphics[width=0.5\textwidth]{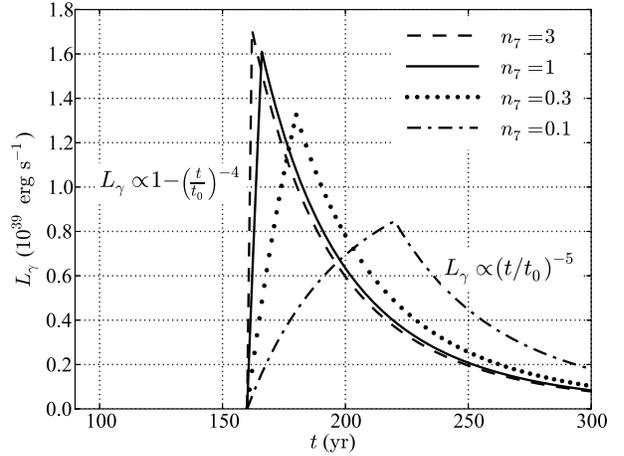}
\caption{Light curves of the TDE afterglows, calculated using the following
parameters, $(\epsilon_{-1},\eta_{-1},m_6,m_*,D_1)=(1,1,1,1,1)$.  Different
lines refer to MCs of different densities. In each calculation, the TDE happens
at the time $t=0$ and the UDS collides with the MC at the time $t_0\simeq160$
yr. Immediately after a collision, the $\gamma$-ray luminosity increases with
time as $1-(t/t_0)^{-4}$ because the newly-produced CRs do not have time to
cool down and hence accumulate inside the MC. On a time-scale longer than
$\tau_{pp}$ the luminosity drops as $(t/t_0)^{-5}$, because the
injection of CRs is now relatively slow so that $L_\gamma$ is limited by
the energy-injection rate [Equation~(\ref{eq:dEkdt})].
\label{fig:lc}} 
\end{figure}

Figure~\ref{fig:lc} illustrates the evolution of $L_\gamma$ as we have just
described. We can see that (i) the $\gamma$-ray emission appears hundreds of
years (depending on $t_0$) after the initial TDE, and (ii) the luminosity
evolves on a time-scale of several decades (also depend on $t_0$).  This
behaviour is very different from that of a tidal flare, which already appears
days after the moment of stellar disruption and lasts at most a couple of years
\citep[e.g.][]{rees88,komossa15}. To highlight this difference, in the
following we will refer to the $\gamma$-ray signature resulting from the UDS-MC
collision as the ``afterglow'' of TDE. 

Figure~\ref{fig:lc} also shows that for a large range of $n_H$, the peak
luminosity of the TDE afterglow is more or less constant, about $10^{39}~{\rm
erg~s^{-1}}$. This result can be understood as a compromise between a
higher $\pi^0$ production rate and a shorter CR cooling time when $n_H$
increases.  A more stringent proof of this insensitiveness to $n_H$ can be
performed by evaluating $L_\gamma$ at the moment $t=t_0+\tau_{pp}$.  Noticing
that $\tau_{pp}\ll t_0$ when $n_H\gg4\times10^5~{\rm cm^{-3}}$, we can re-write
$C(t_0+\tau_{pp})$ as $4\tau_{pp}/t_0$, accurate to first-order, and
consequently Equation~(\ref{eq:Lgamma}) reduces to
\begin{align}
L_{\rm peak}&=(\epsilon\eta/\kappa)4E_k/t_0\label{eq:Lpeak1}\\
&\simeq1.6\times10^{39}~{\rm erg~s^{-1}}
\epsilon_{-1}\eta_{-1}m_6^{1/2}m_*^{2}r_*^{-3/2}D_{1}^{-1},\label{eq:Lpeak2}
\end{align}
where we have normalized $\epsilon$ and $\eta$ using their typical values such
that $\epsilon_{-1}=\epsilon/0.1$ and $\eta_{-1}=\eta/0.1$. It is now clear
that $L_{\rm peak}$ does not depend on $n_H$.

Although insensitive to $n_H$, $L_{\rm peak}$ does depend on other model
parameters.  To see more clearly the dependence, we adopt the relation
$r_*\propto m_*^{0.6}$ for main-sequence stars with $m_*>1$ (see references in
\S\ref{sec:initial}) and derive $t_0\propto m_6^{-1/6}m_*^{0.03}D_1$ and
$L_\gamma\propto m_6^{1/2}m_*^{1.1}D_1^{-1}$.  These scaling relations indicate
that a closer MC, a more massive SMBH or tidal disruption of a more massive
star generally leads to a brighter afterglow, but the time delay between the
initial tidal flare and the later afterglow depends only on the location of the
MC.

\subsection{Spectrum}\label{sec:spec}

From an observational point of view, it is important to understand the
spectral energy distribution of a $\gamma$-ray source, because a $\gamma$-ray
telescope is sensitive to only a limited energy band of photons.

To calculate the $\gamma$-ray spectrum of our TDE afterglow, we use the public
code \texttt{LibppGam} \citep{kafexhiu14}.  As an input to the code a CR
spectrum is needed, and we use Equation~(\ref{eq:CRspec}) to compute it. We set
the minimum energy of the CRs to $0.3$ GeV, which is the $\gamma$-ray kinetic
limit \citep{kafexhiu14}, and set the maximum energy to $T_{p,{\rm max}}= 100$
TeV, above which the CRs would escape from a MC of our interest (see
\S\ref{sec:cr}).  For illustrative purposes, we adopt $\epsilon
E_kC(t_0+\tau_{pp})$ as the total energy of CRs inside MC, where $\epsilon$ is
fixed at our fiducial value $0.1$.  The corresponding total luminosity is
equivalent to $L_{\rm peak}$.  It is worth mentioning that \texttt{LibppGam}
calculates the CR-to-pion conversion efficiency based on particle physics,
therefore, there is no need to specify $\eta$ in the calculation. 

\begin{figure}
\includegraphics[width=0.5\textwidth]{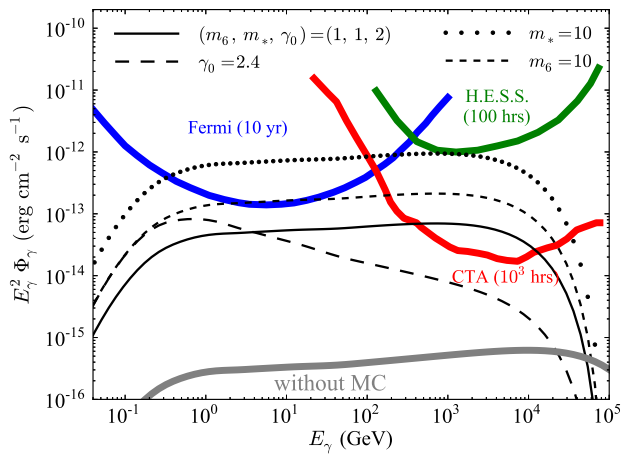}
\caption{The $\gamma$-ray spectra of four representative TDE afterglows (thin
black lines) locating at the same distance of $5$ Mpc. The model parameters, by
default, are $(\epsilon_{-1},m_6,m_*,D_1,\gamma_0)=(1,1,1,1,2)$
unless mentioned otherwise in the legend.  For comparison, the thick grey line
shows the typical $\gamma$-ray spectrum in the case without MCs, where an UDS
interacts only with a smooth ISM of a density of $n_H=10^4~{\rm cm^{-3}}$ (see
\S\ref{sec:noMC} for details).  The other three thick lines with blue, green
and red colors are the sensitivity curves for, respective,
Fermi-LAT, H.E.S.S. and CTA.\label{fig:5Mpc}} 
\end{figure}

Figure~\ref{fig:5Mpc} shows four representative spectra (thin black curves)
calculated in the above way.  By default, the model parameters are
$(\epsilon_{-1},m_6,m_*,D_1,n_7,\gamma_0)=(1,1,1,1,1,2)$ unless indicated
otherwise. In these calculations, we have assumed a distance of $5$ Mpc for the
source.  For comparison, the distance of the Andromeda Galaxy (M31) is about
$0.8$ Mpc, and the diameter of the Local Group is about $3$ Mpc.  A more
thorough study of the effect of distance will be presented in the next section. 

Figure~\ref{fig:5Mpc} shows that when $\gamma_0=2$ the $\gamma$-ray spectrum is
relatively flat.  This flatness reflects the fact that $\eta$ is almost
constant in a large range of energy band. This constancy also causes the
steepening of the $\gamma$-ray spectrum when $\gamma_0$ increases to $2.4$.  In
the following we will focus on the case $\gamma_0=2$ (the universal power law),
motivated by the fact that the shock in our system has an extremely large Mach
number (\S\ref{sec:shock}).

\section{Detectability}\label{sec:detect}

To understand whether a TDE afterglow is detectable, we need first to know our
instruments. Therefore, we consider here three $\gamma$-ray telescopes, namely,
Fermi-LAT, H.E.S.S. and CTA, as representatives of the technology at present
and in the near future.

We say that a telescope can ``resolve'' a TDE afterglow if part of the
$\gamma$-ray spectrum is above the sensitivity curve of the telescope. This
afterglow mostly likely will look like a point source if it is extragalactic given a typical angular resolution of $1$ arcmin for a
$\gamma$-ray telescope \citep[e.g. see][for CTA]{funk13}.  Otherwise, if the
entire spectrum is below the lowest point of the sensitivity curve, the object
is ``unresolvable'', and will only contribute to the diffuse extragalactic
$\gamma$-ray background \citep{ackermann15b} in the field-of-view of the
telescope. 
 
The sensitivity curves of the three telescopes are shown in
Figure~\ref{fig:5Mpc}.  For H.E.S.S. and CTA, the sensitivity curves are
adopted from \citet[][$5\sigma$ detection]{funk13}, and for Fermi-LAT from its
performance webpage\footnote{{\scriptsize
http://www.slac.stanford.edu/exp/glast/groups/canda/lat\_Performance.htm}}. It
is clear that CTA, due to its planned superior sensitivity, can resolve all the
representative TDE afterglows except the one with $\gamma_0=2.4$; it is best
suited to search for the TDE afterglows.

From now on,  we will quantify the number $N_\gamma$ of TDE afterglows
resolvable by each of the three telescopes.  At the end of this section, it
will become clear that $N_\gamma$ is small for Fermi-LAT and H.E.S.S., but is
large enough for CTA to make a detection likely. This is the reason that so far
neither Fermi-LAT nor H.E.S.S. has make clear detection of the TDE afterglows.

We will restrict the following analysis to a single population of TDE
afterglows with the same SMBH masses ($m_6$), stars ($m_*$) and MCs ($D_1$).  This
simplification enables us to derive basic results that can be generalized in
the future to account for multiple populations of afterglows. The
generalization, however, requires knowledge that are currently unavailable,
such as the space density of low-mass SMBHs
\citep[$M_\bullet\sim10^6~M_\odot$,][]{stone14} and the spatial distribution of
MCs in other galaxies. For this reason, we have to postpone it to a future
work.  In this paper, we will circumvent these uncertainties by assigning a
typical value to each of the parameters of $(m_6,m_*,D_1)$, based on the
current understanding of TDEs and MCs.  

Given the above restriction, we derive $N_\gamma$ in three steps.  (i)~Estimate
the maximum distance $r_\gamma(t)$ within which a source can be resolved by a
given telescope.  This maximum distance is a function of $t$ (same definition
as before and we consider only $t>t_0$) because the $\gamma$-ray luminosity
decays with time as $L_\gamma(t)\simeq L_{\rm peak} (t/t_0)^{-5}$.  Since we
are mostly interested in the case of $n_H>10^6~{\rm cm^{-3}}$, we find that the
rising phase (with a duration of $\tau_{pp}$) before the luminosity peak is
relatively short ($\tau_{pp}\ll t_0$ and also see Figure~\ref{fig:lc}), so we
have neglected its contribution to $N_\gamma$.  (ii)~Use $r_\gamma(t)$ to
derive a detectable volume $V_\gamma(t)=4\pi r_\gamma^3(t)/3$, and estimate the
number $\Delta N_\gamma(t)$ of TDE afterglows that are inside this volume and
are in the same time interval of $(t,t+\Delta t)$.  (iii)~Sum up the numbers
$\Delta N_\gamma(t)$ in all intervals at $t>t_0$ to get $N_\gamma$. 

We now address step (i). We know that $r_\gamma(t)$ is proportional to
$L_\gamma^{1/2}(t)$, and therefore to $L_{\rm peak}^{1/2}(t/t_0)^{-5/2}$. Now
substituting $L_{\rm peak}$ with Equation~(\ref{eq:Lpeak2}) and introducing a
normalization factor $r_0$ that will be determined later, we derive
\begin{equation}
r_\gamma(t)\simeq r_0~\epsilon_{-1}^{1/2}m_6^{1/4}m_*^{0.55}D_1^{-1/2}(t/t_0)^{-5/2}\label{eq:rgamma}.
\end{equation}
We note that $r(t)$ does not depend on $n_H$ because neither $L_{\rm peak}$ nor
$t_0$ depends on it.

\begin{figure} \includegraphics[width=0.5\textwidth]{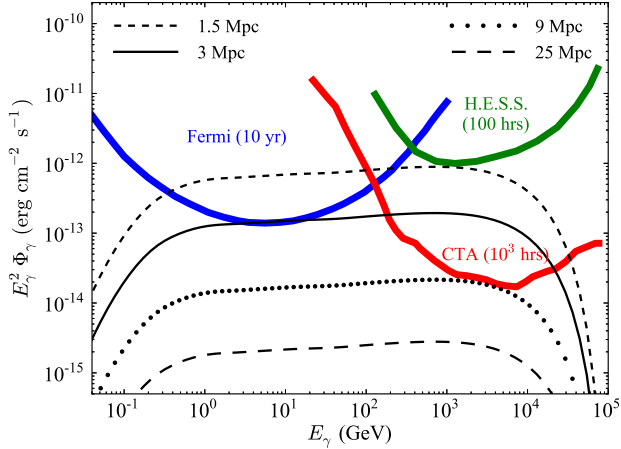} \caption{The
spectra (thin black lines) of a TDE afterglow with
$(\epsilon_{-1},m_6,m_*,D_1,\gamma_0)=(1,1,1,1,2)$ at four different distances.
The thick colored lines are the same telescope sensitivity curves as in
Figure~\ref{fig:5Mpc}.  \label{fig:dis}} \end{figure}

To quantify $r_0$,  we recall that $r_0=r_\gamma(t_0)$ if we adopt the
following parameters for our model $(\epsilon_{-1},m_6,m_*,D_1)=(1,1,1,1)$.
Therefore, $r_0$ is the maximum distance within which a TDE afterglow with the
above fiducial parameters can be resolved.  We can find $r_0$ in
Figure~\ref{fig:dis}, which shows four spectra of the fiducial TDE afterglow at
four different distances, assuming $t=t_0$ in the calculation.  A comparison of
these spectra with the sensitivity curves of the telescopes suggests that
$r_0\simeq3$, $1.5$ and $9$ Mpc for Fermi-LAT, H.E.S.S. and CTA.  Therefore, we
find again that  CTA is the most powerful among the instruments considered here
to search for the TDE afterglows.

In particular, using $r_0=9$ Mpc for CTA and the scaling relation
$r_\gamma(t_0)\propto m_6^{1/4}$ presented in Equation~(\ref{eq:rgamma}), we
find that $r_\gamma(t_0)\simeq16$ Mpc if $m_6$ increases to $10$, which is the
most probable black-hole mass according to TDE observations \citep{stone14}.
This detectable range already reaches the Virgo Cluster.
 
Knowing $r_\gamma(t)$, and therefore $V_\gamma(t)$, we can move on to step
(ii). Since TDEs are a random process in time, we can calculate $\Delta
N_\gamma(t)$ using the formula $\Delta N_\gamma(t)=f_c\Gamma V_\gamma(t)\Delta
t$. The factor $f_c$ denotes the fraction of the sky covered by MCs when viewed
from a central SMBH. This covering factor determines the probability of an UDS
to collide with a MC. In the following we adopt $40\%$ as our fiducial value
(see \S\ref{sec:intro}).  The other parameter $\Gamma$ is the volumetric TDE
rate, an observable usually given in unit of ${\rm Mpc^{-3}~yr^{-1}}$. Later in
this section we will discuss the typical value of $\Gamma$.

In step (iii) we need to sum up all the $\Delta N_\gamma(t)$ for $t> t_0$.  In
the limit $\Delta t\ll t_0$, the summation is equivalent to the integration
$N_\gamma\simeq f_c\Gamma\int_{t_0}^{\infty}V_\gamma(t)dt$, which leads to
\begin{equation}
N_\gamma\simeq12\epsilon_{-1}^{3/2}m_*^{1.6}D_1^{-1/2}\Gamma_{-4}
\left(\frac{f_c}{0.4}\right)
\left(\frac{m_6}{10}\right)^{7/12}
\left(\frac{r_0}{9~{\rm Mpc}}\right)^3,\label{eq:Ngamma}
\end{equation}
where $\Gamma_{-4}\equiv\Gamma/(10^{-4}~{\rm Mpc^{-3}~yr^{-1}})$. In
Equation~(\ref{eq:Ngamma}) we have normalized $m_6$ by $10$ because, as has
been mentioned before, it is the most typical value for the TDEs detected so
far \citep{stone14}.  

For $\Gamma$, a series of observational and theoretical studies have derived a
wide range of values, between $10^{-7}$ and $10^{-4}~{\rm Mpc^{-3}~yr^{-1}}$
\citep[e.g.][]{magorrian99,donley02,wang04,esquej08,gezari09,vanvelzen14,sun15,holoien16,stone14}.
It is important to point out that these earlier works were studying TDEs at
relatively large distances, in a redshift range of $0.01\la z\la1$
\citep{komossa15}, so for their purposes the space density of SMBHs averages
out to about ${\cal O}(10^{-2})~{\rm Mpc^{-3}}$
\citep[e.g.][]{donley02,stone14}.  We, however, are interested mainly in nearby
SMBHs, because $r_0$ is relatively small. In the nearby universe there is
apparently an overdensity of SMBHs -- we already know three within a distance
of $1$ Mpc \citep[the MW, M31 and M32, and see][for more SMBHs in the Local
Group]{kormendy13}.  To account for this overdensity, we will consider
$\Gamma=10^{-4}~{\rm Mpc^{-3}~yr^{-1}}$ as our fiducial value, which
corresponds to $\Gamma_{-4}=1$.

Knowing the value of $\Gamma$, we now return to Equation~(\ref{eq:Ngamma}). A
CTA-like telescope corresponds to $r_0=9$ Mpc, so it could resolve about $12$
TDE afterglows in all sky as point sources.  This result also suggests that in
order to detect ${\cal O}(10^2)$ TDE afterglows, a telescope four times more
sensitive than CTA is needed.  On the other hand, for Fermi-LAT $N_\gamma$
would be about $27$ times smaller, i.e $N_\gamma<0.5$, because $r_0\simeq3$ Mpc
is three times smaller than before. The number $N_\gamma$ for H.E.S.S. would be
eight times even smaller because of the smaller $r_0$.

Note that these numbers do not include the contributions from dwarf galaxies.
Later in \S\ref{sec:dis} we will discuss the TDE afterglows in dwarf galaxies.
 
\section{Without MCs}\label{sec:noMC}

It is important to know what would happen if there were no MCs around SMBHs.
In this case, neutral pions and $\gamma$-rays could still be generated by the
interaction of UDSs with a smooth ISM. For the GC, earlier calculations showed
that the corresponding $\gamma$-ray luminosity is as high as
$10^{39}-10^{40}~{\rm erg~s^{-1}}$ during the beginning of the interaction
\citep{cheng06,cheng07,cheng11,cheng12}.  Here we will revisit the calculation
using the updated UDS model.

For the sake of comparison, we adopt the same ISM model as in the earlier
studies -- the ISM has a constant proton density of $n_H\sim10^4~{\rm cm^{-3}}$
\citep{khokhlov96,cheng06}. We will discuss the dependence of our results on
$n_H$ later in this section.  Given this density, the characteristic length to
stop UDS from free streaming, i.e.  $\Delta D$ in Equation~(\ref{eq:DeltaD}),
increases to about $\Delta D'\simeq2$ pc. Consequently, the Sedov-like
expansion, during which most CRs are produced, starts at about
$t'_0\simeq\Delta D'/v_0\simeq300$ yr.

Since $t'_0$ is much shorter than the proton cooling time-scale in the ISM,
which is now about $\tau'_{pp}\simeq5900$ yr according to
Equation~(\ref{eq:taupp}), we are allowed to treat the CR-injection process as
an instantaneous event, and separate it from the later CR-cooling process.  For
this reason, we take $E'_{\rm CR}=\epsilon E_k$ as the total energy of CRs
eventually injected into the ISM. 

Knowing $n_H$ and $E'_{\rm CR}$, we can calculate the $\gamma$-ray luminosity
with
\begin{align} 
L'_\gamma&=\eta(\sigma_{pp}n_Hc) {E}_{\rm CR}\\
&\simeq1.2\times10^{37}~{\rm erg~s^{-1}}
\times\epsilon_{-1}\eta_{-1}m_6^{1/3}m_*^{1.1}n_4,\label{eq:Lism}
\end{align}
where $n_4=n_H/(10^4~{\rm cm^{-3}})$. Unlike that in the case of UDS-MC
interaction [Equation~(\ref{eq:Lpeak2})], the luminosity here depends on the
density $n_H$, because now it is limited not by the injection rate of CRs but
the cooling rate.  The evolution of the luminosity is also different: It rises
on a time-scale of $300$ years ($t'_0$) to reach $L'_\gamma$ and then persists
for about $6000$ years ($\tau'_{pp}$). To distinguish this $\gamma$-ray
signature from the previous TDE afterglow, we call it the second-type
afterglow.

As for the $\gamma$-ray spectrum, we calculate it following the same procedure
described in \S\ref{sec:spec}, except for three modifications: (i) The total
CR energy is now $E'_{\rm CR}$, (ii) the proton density in the background is
set to $10^4~{\rm cm^{-3}}$ and (iii) the maximum CR energy is $T_{p,{\rm
max}}=1$ PeV, adopted from \cite{cheng12}. The result is shown in
Figure~\ref{fig:5Mpc} as the thick grey line. It clearly shows that without
MCs, the $\gamma$-ray afterglow would be much fainter.

We note that the luminosity given by Equation~(\ref{eq:Lism}) is more than
$10^2$ times smaller than the peak luminosity $10^{39}-10^{40}~{\rm
erg~s^{-1}}$ derived by \cite{cheng06,cheng07}, who considered a similar
scenario of $\gamma$-ray emission following a TDE.  This disparity is caused
mainly by the much greater $E'_{\rm CR}$ used by Cheng et al. in their model,
which is about $6\times10^{52}~{\rm erg}$. Such a large energy is unlikely to
be provided by an UDS \citep{guillochon15b}, but a jet, whose formation may
require special conditions not satisfied by most TDEs \citep{decolle12}.
 
Since the luminosity $L'_\gamma$ is about $10^2$ times lower than $L_{\rm
peak}$ in \S\ref{sec:L}, we expect the maximum distance at which a source is
resolvable, $r'_\gamma$, to be $10$ times smaller and the corresponding volume,
$V'_\gamma$, to be $10^3$ times smaller. These considerations lead to
$r'_\gamma\simeq
0.9~\epsilon_{-1}^{1/2}\eta_{-1}^{1/2}m_6^{1/6}m_*^{0.55}n_4^{1/2}$ Mpc for
CTA.  Correspondingly, the detectable volume is $V'_\gamma=4\pi
{r'}_\gamma^3/3$.  Now neither $r'_\gamma$ nor $V'_\gamma$ depend on time,
because the $\gamma$-ray emission that we are considering now lasts from the
time $t=t'_0$ to $t'_0+\tau'_{pp}$ during which $L'_\gamma$ is mostly constant.

Knowing the detection volume $V'_\gamma$ and the duration $\tau'_{pp}$ of the
second type of TDE afterglows, we can calculate the number of resolvable point
sources $N'_\gamma$ using the formula $\tau'_{pp}\Gamma V'_\gamma$. Note that
the covering factor $f_c$ does not appear here, because we assumed that ISM
distributes uniformly around a SMBH.  With the parameters relevant to CTA, we
derive that
\begin{equation}
N'_\gamma\simeq \tau'_{pp}\Gamma V'_\gamma
\sim6\epsilon_{-1}^{3/2}\eta_{-1}^{3/2}m_*^{1.6}n_4^{1/2}\Gamma_{-4}\left(\frac{m_6}{10}\right)^{1/2}.
\end{equation}
We expect these CTA sources to be relatively close to the earth, because
$r'_\gamma\simeq1.5(m_6/10)^{1/6}$ Mpc when SMBHs with $m_6=10$ are considered.
For the same reason given at the end of \S\ref{sec:detect}, the number of point
sources resolvable by Fermi-LAT and H.E.S.S.  is negligibly small. 

In the above calculation, we have assumed $n_H=10^{4}~{\rm cm^{-3}}$ for the
sake of comparison. However, for the GC, theoretical models of the accretion
flow onto the SMBH (Sgr A*) suggested that the ISM density gets as high as
$10^{4}~{\rm cm^{-3}}$ only within the central $10^{-3}$ pc \citep{yuan03}.
Moreover, observations of hot and X-ray-emitting ISM within a distance of $1$
pc from Sgr A* showed that the ISM density is only ${\cal O}(10^{2})~{\rm
cm^{-3}}$ on average \citep{baganoff03}. If the circum-nuclear media in
other galaxies are similar to the ISM in the GC \citep{generozov15}, we would
have a much lower value for $n_H$ than what has been assumed above.
Correspondingly, both $L'_\gamma$ and $N'_\gamma$ would be lowered, making it
more difficult to detect the second type of TDE afterglows.

\section{Discussions}\label{sec:dis}

In this paper we have investigated a scenario likely to happen following a
large number of TDEs (\S\ref{sec:intro} and \S\ref{sec:MC}), in which the
ejecta, or ``UDS'' (\S\ref{sec:initial}), launched from a disrupted star
travels to a distance of ${\cal O}(1)$ pc and collides into a MC.  We have
shown that the collision, which happens hundreds of years after the initial
TDE, would produce a strong shock into the MC (\S\ref{sec:shock}), giving rise
to a large amount of CRs with a kinetic energy as high as $1$ PeV
(\S\ref{sec:cr}).  These CRs, after escaping from the shock region, would
collide with the non-relativistic protons in the MC, producing $\pi^0$ and
subsequently $\gamma$-rays in a wide energy range of $0.1\sim10^5$ GeV
(\S\ref{sec:gammaray}).  According to our calculation, this $\gamma$-ray
signature, which we call ``TDE afterglow'', would be detectable by a CTA-like
telescope out to a distance of $10-20$ Mpc (\S\ref{sec:detect}).  On the other
hand, if there were no MCs around SMBHs, there would be a second type of
$\gamma$-ray afterglows resulting from the interaction of UDSs with a smooth
ISM, which are much fainter than the first type (\S\ref{sec:noMC}).

So far we have not considered IMBHs ($10^3\le M_\bullet/M_\odot<10^6$), which
exist probably in the nuclei of many dwarf galaxies \citep{kormendy13}.  They
produce TDE afterglows in the same way as SMBHs do, except that the typical
luminosity is different. When an IMBH, $M_\bullet=10^3~M_\odot$ for example,
tidally disrupts a main-sequence star, the peak luminosity of the $\gamma$-ray
afterglow would be $5\times10^{37}~{\rm erg~s^{-1}}$ according to
Equation~(\ref{eq:Lpeak2}) (with MCs), or about $10^{36}n_4~{\rm erg~s^{-1}}$
according to Equation~(\ref{eq:Lism}) (without MCs). On the other hand, IMBHs
are also able to tidally disrupt white dwarfs. If such an event happens, the
kinetic energy of the UDS according to Equation~(\ref{eq:Ek}) could
significantly exceed $10^{50}$ erg, because a white dwarf is $10^2$ times
smaller than the sun. As a result, the $\gamma$-ray luminosity would be much
higher. 

Since dwarf galaxies have been found in large amount in the Local Universe, the
number of TDE afterglows would also be large if IMBHs are ubiquitous in dwarf
galaxies.  Intriguingly,  \citet{GS15,hooper15} recently reported a possible
detection of $\gamma$-ray excess, about $5\times10^{33}~{\rm erg~s^{-1}}$ in
the Fermi-LAT band, in a nearby dwarf galaxy Renticulum II \citep[but see][for
a different result]{drlica-wagner15}. This result immediately caused a debate
about the origin of this $\gamma$-ray source, including dark-matter
annihilation \citep[e.g. see][for an earlier proposal]{diemand08}. Our scenario
of TDE afterglow can also explain the $\gamma$-ray luminosity of Renticulum II.
For example, if a TDE with $M_\bullet=10^3~M_\odot$ and $m_*=1$ happened $3200$
years ago, then after $500$ years the UDS would have traversed a distance of
$1$ pc to hit a MC there (we find $t_0\simeq 500D_1$ yr in this case), and
today the $\gamma$-ray afterglow could have decayed to a luminosity of
$5\times10^{33}~{\rm erg~s^{-1}}$.

Since dark-matter annihilation, among with other sources such as star-bursts
and a population of millisecond pulsars (MSPs), could also produce
$\gamma$-rays above $1$ GeV in galaxy centres
\citep{su10,abazajian14,brandt15b}, it is critical to find a way of separating
these sources from our TDE afterglows.  

We find that the $\gamma$-ray luminosities from dark-matter annihilation,
probably about ${\cal O}(10^{37})~{\rm erg~s^{-1}}$ for a MW-like galaxy
\citep{fermi15}, and from MSPs, on average $10^{34}-10^{35}~{\rm erg~s^{-1}}$
for one MSP \citep{hooper15b}, are both much lower than the peak luminosity of
the first-type TDE afterglows, which, we now know, is about $10^{39}~{\rm
erg~s^{-1}}$. This difference in luminosity could be used to distinguish TDE
afterglows from the other two $\gamma$-ray sources. The limitation of this
method is that it is effective for only ``young'' TDE afterglows.  Because by
the time of $t=(100)^{1/5}t_0\simeq2.5t_0$, the luminosity of the afterglow
would have dropped to $1\%$ of its peak value, and it is no longer
distinguishable from dark-matter-annihilation signal or a population of ${\cal
O}(10^2)$ MSPs.

On the other hand, some star-forming galaxies could be as luminous as
$10^{39}~{\rm erg~s^{-1}}$ in $\gamma$-ray \citep{ackermann12}, and in this
case a method other than comparing luminosities is needed.  In the following we
outline three methods that could be useful to separate the first type of TDE
afterglows (UDS-MC-induced) from star-forming galaxies. 

\begin{enumerate}

\item As we have seen in Figure~\ref{fig:lc}, a TDE afterglow could be highly
variable on a time-scale of ten years, especially at the beginning.  On the
other hand, the $\gamma$-ray emission from a star-burst (or a population of
pulsars and dark-matter annihilation) is normally constant for (at least)
million of years.  Therefore, a variable $\gamma$-ray source with the
characteristic luminosity ${\cal O}(10^{39})~{\rm erg~s^{-1}}$ and spectral
index $\gamma_0=2$ would be a smoking-gun evidence for the first type of TDE
afterglow.

\item In \S\ref{sec:cr} we found that there is a critical kinetic energy,
$T_p\sim10n_7^2$ TeV, above which the CRs are able to escape from a MC before
significantly losing their energies. Because of the existence of this critical
energy, we expect a discontinuity visible in the $\gamma$-ray spectrum at a
photon energy comparable to this critical energy -- the intensity of the
spectrum above this critical photon energy would be much weaker than that
below.  Such a ``spectral kink'' lies in the sensitive window of CTA, and if
detected, could provide another evidence for our first type of TDE afterglow.
To better understand the time dependence, spectral shape and contrast of this
discontinuity, it will be helpful to simulate the propagation of CRs inside a
dense MC ($n_H\sim10^7~{\rm cm^{-3}}$).

\item 
Observations of star-formation galaxies by Fermi-LAT revealed a tight
correlation between the $\gamma$-ray luminosity ($L_\gamma$ in the band
$0.1-100$ GeV) and the luminosity in far-infrared \citep[$L_{\rm FIR}$,
see][]{ackermann12}.  This correlation likely reflects a fundamental
relationship between the star-formation rate (SFR) and the rate of CR
production.  According to this relationship, $L_\gamma$ decreases linearly from
about $10^{39}~{\rm erg~s^{-1}}$ when the SFR is about $1~M_\odot~{\rm
yr^{-1}}$ (like the MW), to about  $10^{37}~{\rm erg~s^{-1}}$ where the SFR is
about $10^{-2}~M_\odot~{\rm yr^{-1}}$ (such as the Small Magellanic Cloud).
Correspondingly, $L_{\rm FIR}$ decreases from about $2\times10^{43}~{\rm
erg~s^{-1}}$ to $2\times10^{41}~{\rm erg~s^{-1}}$.  Based on these
luminosities, it can be seen that if a TDE afterglow with
$L_\gamma\sim10^{39}~{\rm erg~s^{-1}}$ appears in a galaxy whose SFR is lower
than $1~M_\odot~{\rm yr^{-1}}$, $L_\gamma$ in total would rise to above
$10^{39}~{\rm erg~s^{-1}}$ but $L_{\rm FIR}$ is unaffected by the TDE
afterglow, since the $L_{\rm FIR}$ of a TDE afterglow cannot exceed
$10^{41}~{\rm erg~s^{-1}}$ according to the cooling rate calculated in
\S\ref{sec:shock}. As a result, a galaxy hosting a TDE afterglow would appear
as an outlier in the $L_\gamma-L_{\rm FIR}$ diagram  relative to the population
of star-forming galaxies, if the SFR of this galaxy is below $1~M_\odot~{\rm
yr^{-1}}$.

\end{enumerate}

If CTA or more sensitive telescopes in the future do not detect any TDE
afterglow, the non-detection, according to Equation~(\ref{eq:Ngamma}), may be
due to one or a combination of the following factors. (i) There is a deficiency
of MCs in other galaxies so that $f_c$ is much smaller than $0.4$.  (ii) MCs on
average lie much further away from SMBHs than a distance of $1$ pc.  (iii) The
volumetric TDE rate $\Gamma$ in the local universe is much lower than
$10^{-4}~{\rm Mpc^{-3}~yr^{-1}}$. (iv) Or the current theory of the UDS from a
TDE may be incomplete.  These knowledges would help us better understand TDEs,
as well as the stellar and gaseous environments around the SMBHs or IMBHs in
normal galaxies.

\section*{Acknowledgments}

We thank Fukun Liu for pointing out the importance of tidal disruption of white
dwarfs.  XC is supported by CONICYT-Chile through Anillo (ACT1101). J.~G. is
supported by the NASA Einstein grant PF3-140108.  GAGV is supported by Conicyt
Anillo grant ACT1102, the Spanish MINECO’s Consolider-Ingenio 2010 Programme
under grant MultiDark CSD2009-00064 and also partly by MINECO under grant
FPA2012-34694.  Part of this work is carried out at the Astronomy Department of
Peking University, supported by the ``VRI concurso estad\'{i}as en el
extranjero'' of PUC.

\end{document}